\documentclass[12pt]{article}
\usepackage{epsf}
\usepackage{graphicx}
\usepackage{natbib}
\usepackage{journals}   

\begin{document}

\title{\vskip -2em \underline{\small Astronomy Reports, 2020, Vol. 64, No. 2, pp. 168--188. DOI: 10.1134/S1063772920020043} \\[2em]
Statistical analysis of the results of 20 years of activity\\ of the International VLBI Service for Geodesy\\ and Astrometry}
\author{Zinovy Malkin$^{1,2}$ \\ (1) Pulkovo Observatory, St. Petersburg 196140, Russia\\ (2) Kazan Federal University, Kazan 420000, Russia\\
        e-mail: malkin@gaoran.ru}
\date{\vskip -2em}

\maketitle

\begin{abstract}
2019 marked the 20th anniversary of the International VLBI Service for Geodesy and Astrometry (IVS). 
This service is the largest and most authoritative organization that coordinates international activities in radio astrometry and VLBI sub-system of space geodesy.
Currently, about 60 antennas located in many countries on all continents participate in the IVS observing programs.
The IVS Data Centers have accumulated more than 18 million observations obtained during more than 17,000 sessions, 
including more than 10,000 Intensive sessions for rapid determination of Universal Time.
The paper traces the dynamics of IVS development based on statistical processing of the array of observations
collected in the IVS Data Centers for the period of 1979--2018.
Various statistics by the years, stations, baselines, and radio sources are provided. 
The evolution of the IVS observational data and the accuracy of results obtained from processing VLBI observations is considered.
\end{abstract}


\section{Introduction}

One of the main modern methods of astronomical research is Very Long Baseline Interferometry (VLBI).
The accuracy of this method has now reached a microarcsecond level, which is implemented in works in the fields of astrophysics
(the size and structure of celestial objects), astrometry (absolute and relative positions of radio sources), stellar astronomy
(parallaxes and motion of Galactic masers, on which, for example, the most accurate studies of the rotation of the Galaxy are based),
celestial mechanics (dynamics and ephemerides of Solar system bodies, including observations of spacecraft and radio beacons).
Astrometric applications of VLBI include various scientific and applied problems in such areas as Earth rotation, geophysics,
Earth's crust deformations and movements, geodesy, and navigation.
This method also plays an extremely important role in establishing the celestial and terrestrial coordinate systems and frames.

Since the accuracy of VLBI observations depends critically on the number of VLBI antennas used and the distance between them
(baselines length), the most interesting data is obtained from observations on global station networks.
The organization of such observations requires, as a rule, well-established interaction between institutions in different countries.
For this purpose, various coordinating centers are organized in the world, one of which is the International VLBI Service for Geodesy and 
Astrometry\footnote{https://ivscc.gsfc.nasa.gov/} (IVS).
In 2019~this service celebrated its 20th anniversary, which was the primary reason for preparing this article.
 
The present paper is not aimed at detailed consideration of the history of radio interferometry.
Nevertheless, we would like to mention some milestones in the development of this technique.
The first radio interferometer, of the so-called sea-cliff type, was constructed in 1945--1946 in Australia \citep{Sullivan1991}.
It consisted of a single antenna, which observed a direct and reflected by the sea surface radio signal from the Sun near the horizon, and was used to study sunspots.
A little bit later, the first Michelson interferometer \citep{Ryle1946} was constructed at Cambridge, which was at first also used for observations of the Sun.
However, already in the late 1940s--early 1950s, the first observations of radio stars were carried out and their coordinates were determined \citep{Ryle1950,Ryle1955}.
This can be considered the beginning of the radio astrometry era, although the term itself appeared apparently in the early 1970s
(the earliest paper containing the term ``radio astrometry'' found by the author was dated 1973)\footnote{After publication of this paper,
the author found the text of the lecture given by Thomas Clark at the Institute of Applied Astronomy in 2005 \citep{Clark2005}.
In that lecture, Clark quoted his note written to Robert Coates on February 27, 1969, where he wrote: 
``$\ldots$ we hope to measure positions [of radio sources] to similar accuracies -- $\sim$0.001 seconds of arc. These measurements might best be called
$'$Radio Astrometry$'$ ''.
Thus, the term radio astrometry appeared several years earlier than it was assumed in the original version of this paper.}

Thus, the history of radio astrometry dates back about 70 years!

The revolutionary stage in the development of radio interferometry was the method of VLBI.
The main idea of this method is to independently record signals at each antenna along with time marks from a highly stable standard,
with following joint correlation processing of these signals.
The fascinating history of the VLBI method origin is described in \citet{Kellermann1988,Moran1998,Clark2003,Matveenko2007}.
In particular, L.I.~Matveenko recalls the first discussions of the VLBI concept and the possibility of its practical implementation as early as
in February 1962 \citep{Matveenko2007}, although the first official paper describing the basics of the VLBI was published later  \citep{Matveenko1965}.
\citet{Ryan1998} gave a brief history of the development of the VLBI work at NASA, in particular, at the Goddard Space Flight
Center (GSFC), which played a leading role in the establishment of the IVS; the GSFC remains the basis of many key elements of this service.

The target of observations in the first practical VLBI experiments of 1965--1967 was Jupiter \citep{Brown1968,Carr1970}.
In the spring of 1967, the first successful VLBI observations of extragalactic radio sources were carried out in Canada and the USA
\citep{Broten1967a,Broten1967b,Bare1967}, which opened the way to a new stage in the development of the radio astrometry based on VLBI observations.
The application of this method has made it possible to drastically increase (by orders of magnitude) the accuracy of many astrometric
and geophysical data and results, such as the celestial coordinate system, the Earth's rotation speed and polar motion, precession and nutation,
as well as tectonic motion and regional deformations of the Earth's crust.
There is no need to repeat here all the tasks and achievements of radio astrometry, as these are described in a number of review papers,
e.g., \citet{Sovers1998,Zharov2011}, which also include an extensive bibliography.

The accuracy of astronomical and geodetic results obtained by radio astrometry directly depends on the number and location of the antennas
that form the observation network.
The more antennas used and the larger the network size (its latitudinal and longitudinal span), the higher the accuracy of the results.
For this reason, the main IVS observing programs involve stations located in different countries and on different continents, including Antarctica.
In the early years of the VLBI development, single experiments were carried out on the basis of individual agreements between research groups.
As this work expanded in the direction of monitoring the Earth orientation parameters (EOP) and plate tectonics, this required a new level of
organization of observing programs, inter-institutional and international cooperation, as well as financial planning. 
An example of such a new organization was the NASA's long-term Crustal Dynamics Project (CDP) launched in 1979 \citep{Ryan1998}.
Later, national VLBI programs and station networks were developed in other countries, such as the USSR/Russia \citep{Finkelstein1993,Finkelstein2004},
Japan \citep{Amagai1989,Kondo1999,Ogi1999}, China \citep{Qian2004}, \citep{Lovell2013}, and Italy \citep{Stagni2016}.

It should be noted that the 1960s and 1970s were not only the time of the formation of the astrometric and geodetic VLBI,
it was also the time of the origin and development of other new methods of space geodesy, such as laser ranging to the Earth's artificial
satellites (SLR) and the Moon (LLR), and radio-technical observations of the satellites of navigation systems.
The tasks that are solved by these three methods overlap significantly and include the construction of the terrestrial coordinate system
and monitoring the Earth's surface movements, as well as the study of the Earth's rotation.
Therefore, the question soon arose of the coordination of work on space geodesy, including observing programs, unification of astronomical
and geophysical models used in the analysis of observations, and derivation of joint combined solutions for the EOP and terrestrial coordinate system.
To solve these problems, at the XVII General Assembly of the International Union of Geodesy and Geophysics (IUGG) in 1979, an International Commission
for Coordination of Space Techniques for Geodesy and Geodynamics (CSTG) was founded.
A few years later, three CSTG subcommissions were formed: GPS, SLR, and VLBI, which later became the basis of independent
International GPS Service (IGS, later International GNSS Service), ILRS (International Laser Ranging Service), and IVS.

The highlights of the history of the IVS creation can be traced to an interesting archival document selection on the IVS
website\footnote{https://ivscc.gsfc.nasa.gov/about/org/documents/index.html}.
Some documents and correspondence relating to this period are also stored in the author's archive.
These materials may serve as a contribution to a separate study on the history of IVS and radio astrometry, but this is beyond the scope of this paper.
We only note here that the first proposal to create a special international VLBI service was sent out by the CSTG subcommittee on September 18, 1997,
and the first meeting of the IVS Directing Board was held on February 11, 1999.
The official announcement of the IVS launch came on March 1, 1999 \citep{Schlueter2002}.
Thus, the organizational period for the creation of the IVS took approximately a year and a half.
In the following sections, we will track the progress in various areas of radio astrometry based on the IVS data and results.


\section{General statistics of the observations}
\label{sect:obsstat}

The statistical analysis presented in this paper is based on the archive of observations and results stored
in the IVS Data Centers\footnote{ftp://cddis.gsfc.nasa.gov/pub/vlbi/ivsdata/}.
It continues and expands the earlier work \citet{Malkin2004}.

To understand the statistics presented below properly, one should consider the following.
Observations are organized in sessions that generally last 24 hours or 1 hour based on a predefined schedule.
Further, the observations from all telescopes are delivered to correlator, where the observables are determined,
such as the radio-interferometric delays and delay rates, which are further used for scientific analysis.
For various reasons, a correlation response has been obtained for approximately 75\% of the observations \citep{Berube2017}.
The correlated data included in the correlator output files have different quality and are marked with the corresponding quality codes.
Observations marked as low quality are usually not used in further scientific processing.
The analysis of the data used in this paper shows that the proportion of such observations in the correlator output files is 13.6\%.
Finally, some observations are rejected during the scientific analysis.
The number of these depends on various reasons, including subjective ones, and, as a rule, is relatively small.

Therefore, the number of observations made at the stations characterizes the stations activity; the number of correlated observations characterizes the effective observational work
of the VLBI network; and the number of observations used to obtain the final result characterizes the scientific and practical return from the VLBI network.
In this paper, we present statistics of all the observations included in the correlator output files for 1979--2018, which are stored in the IVS Data Centers.
The first VLBI session stored in the IVS archive was observed on August 3--6, 1979, at three U.S. stations: HAYSTACK, NRAO~140, and OVRO~130 (hereinafter, the station names are
used as they are given in the IVS catalog\footnote{https://ivscc.gsfc.nasa.gov/stations/index.html}).
In total, these statistics include 17'382'383 observations of 5453 radio sources performed at 183 stations and 2403 baselines.
The total duration of all sessions was more than 7208 days, which is equivalent to almost 20 years of continuous observation!

The data below refer to the observations in the S/X bands in the Mark-4 standard.
In 2015, an international network of stations operating according to the new technology standard VLBI2010 \citep{Behrend2009} began to work.
At first, it was two stations, GGAO12M (Maryland, USA) and WESTFORD (Massachusetts, USA), on which new equipment and observation technology were tested \citep{Niell2018}.
In 2016--2017, antennas KOKEE12M (Hawaii, USA), WETTZ13S (one of the two new antennas at the Wettzell station, Germany), RAEGYEB (Yebes, Spain), ISHIOKA (Ishioka, Japan),
and ONSA13NE (one of the two new antennas at the Onsala station, Sweden) were added to this network.
With the exception of station WESTFORD, which was created on the basis of the existing 18-meter antenna, new fast antennas with a diameter of 12--13.5 m were installed at other stations.
The network of the new stations was called the VLBI Global Observing System\footnote{https://ivscc.gsfc.nasa.gov/technology/vgos-general.html} (VGOS).
This network performs mostly test observations so far; with few exceptions, they are not contained in the IVS database.
Therefore, they were not taken into account in this work.
However, it can be noted that under this program in 2015--2018, a total of approximately 100 sessions from 1 to 24 hours long were conducted, in which approximately
500'000 observations were obtained.

The main scientific and practical results of radio astrometry observations are obtained at networks that
consist of at least three stations and have a sufficient latitudinal and longitudinal span.
Such observations make it possible to determine the complete set of astronomical and geophysical parameters related to the terrestrial
and celestial coordinate systems, the rotation of the Earth, the Earth's crustal movements, and others.
In particular, they allow one to determine all three types of the EOP: Earth's pole coordinates, celestial pole coordinates, and Universal Time (UT1).
Most of these observations are organized in the form of 24-hour sessions.
However, these observations are resource-consuming; for this reason, the 24-hour sessions for determining the EOP have been carried out in recent
years, two to three times a week on average.
The task of organizing continuous VLBI observations at global networks, including those for determining the EOP, which are extremely important both for studying the Earth's rotation
and for various other users of these data, has been discussed repeatedly \citep{MacMillan1999,Schuh2002,Behrend2009,Nothnagel2016} and is being gradually implemented.
A prototype of such observations is special continuous two-week CONT observational campaigns, which took place once every three years: CONT02 (October 2002, 8 stations,
49'826 observations); CONT05 (September 2005, 11 stations, 96'437 observations); CONT08 (August 2008, 11 stations, 153'738 observations); CONT11 (September 2011, 13 stations,
145'214 observations); CONT14 (May 2014, 17 stations, 287'234 observations); and CONT17 (November-December 2017, 27 stations, 372'573 observations).
The corresponding plots below clearly show jumps in the number of observations in the years of the CONT campaigns, especially the recent ones.

Requirements for the latency of results of different kinds of EOP determined from VLBI observations are different.
The Earth's pole coordinates are determined with high accuracy and efficiency using satellite methods as well, while the VLBI remains
the primary method for determining UT1 and celestial pole movement.
In particular, increasing the rapidness of the UT1 determinations, even with slightly reduced accuracy, is extremely important for many practical applications.
To comply with this request, regular observations called UT1 Intensive were started in April 1984 in the form of short, normally one-hour, sessions
on two stations (one baseline) and less often at several stations.
Observations under the IVS UT1 Intensive programs are carried out almost daily, and sometimes twice a day on different networks.
Work is also underway to increase the processing speed of these data, as a result of which the delay in obtaining UT1 can be greatly reduced,
approaching the delay in the results obtained by satellite methods \citep{Sekido2008,Finkelstein2011}.

Since the purpose and applications of the 24-hour and Intensive sessions are different, separate statistics on these types of observing programs
are given below when appropriate along with the joint statistics based on the whole data set.
To avoid possible ambiguities, the sessions were divided into 24-hour and Intensive sessions as they designated in the IVS observing 
schedule\footnote{https://ivscc.gsfc.nasa.gov/program/operafiles.html}.

\begin{figure*}
\centering
\resizebox{\textwidth}{!}{\includegraphics[clip]{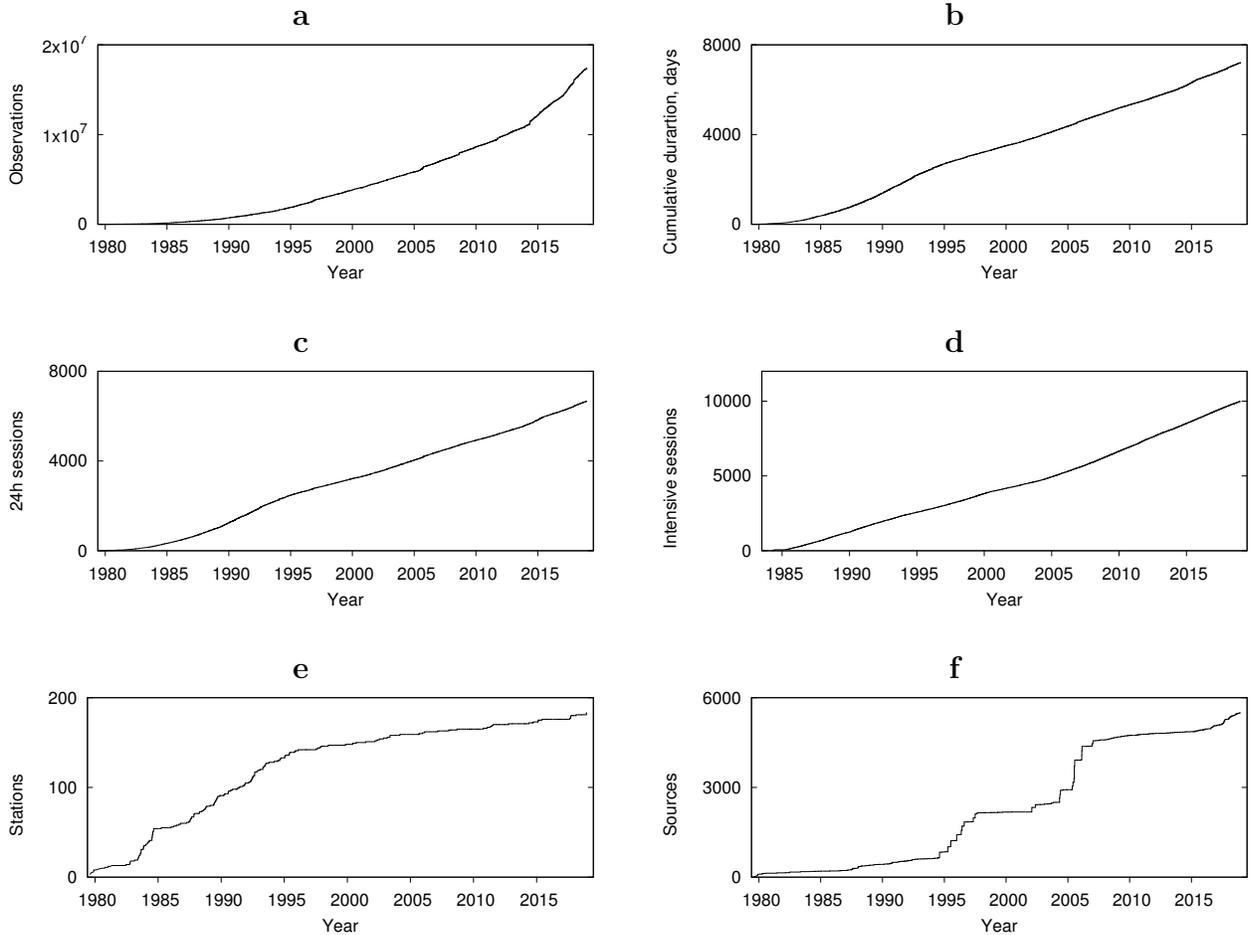}}
\caption{Cumulative statistics of the IVS data: (a) number of observations; (b) total summary duration of the sessions in days; (c) number of 24-hour sessions;
  (d) number of Intensive sessions; (e) number of stations; (f) number of sources.}
\label{fig:stat_cumul}
\end{figure*}

\begin{figure*}
\centering
\resizebox{\textwidth}{!}{\includegraphics[clip]{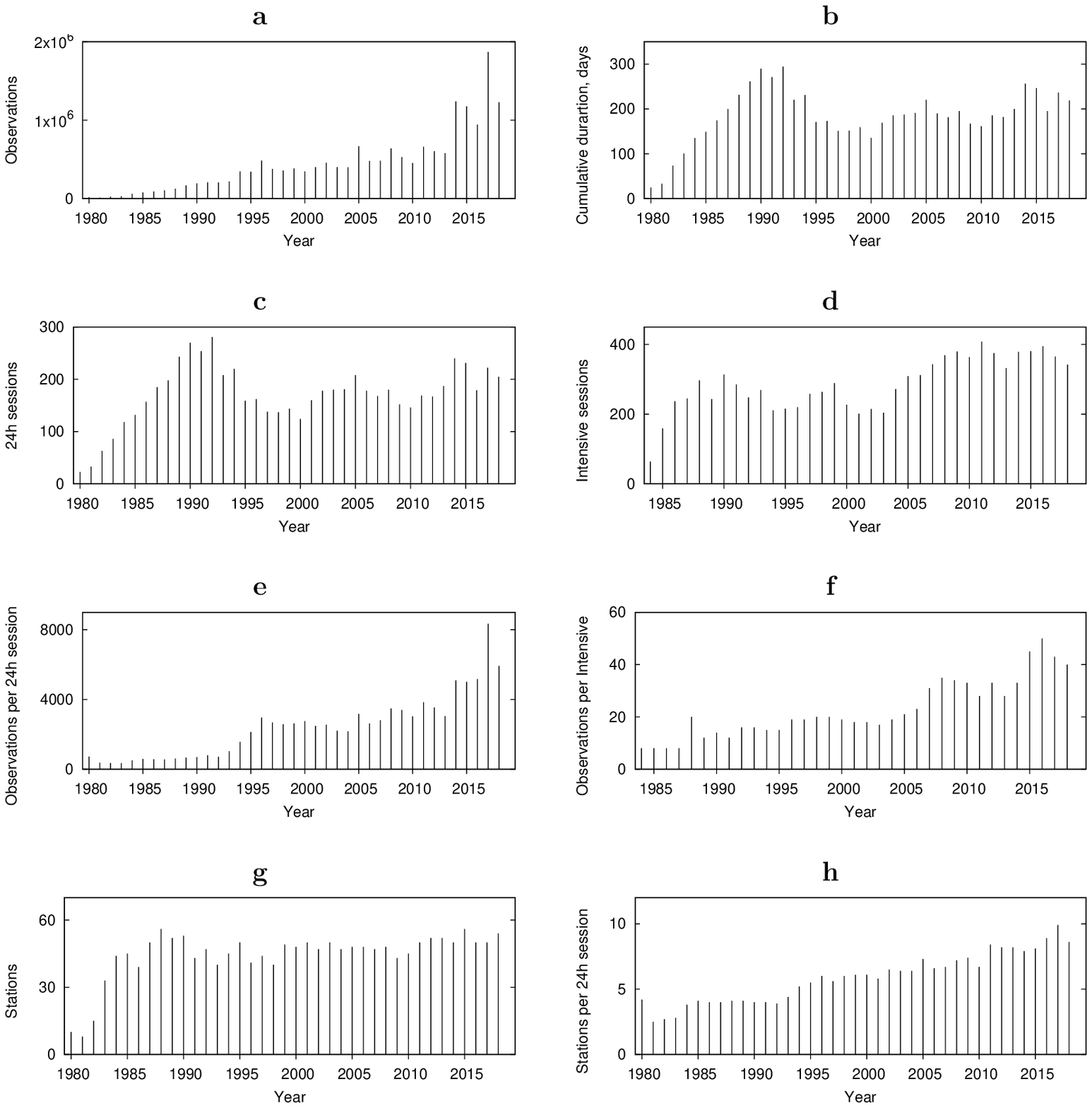}}
\caption{Statistics of the IVS data by years: (a) number of observations; (b) total summary duration of the sessions in days; (c) number of 24-hour sessions; 
  (d) number of Intensive sessions; (e) average number of observations in one 24-hour session; (f) average number of observations in one Intensive session; 
  (g) number of stations; (h) average number of stations participated in one 24-hour session.}
\label{fig:stat_year}
\end{figure*}

Figure~\ref{fig:stat_cumul} shows the growth dynamics of the number of observations, the number of stations and radio sources from 1979 to 2018,
and Fig.~\ref{fig:stat_year} shows similar statistics by year.
The data in Fig.~\ref{fig:stat_cumul} show that the total duration of the observing sessions, the number of 24-hour sessions and the number of Intensive sessions
increases nearly uniformly, while other plots contain some interesting details.
For example, the accelerated increase in the number of observations (Fig.~\ref{fig:stat_cumul}a) corresponds to an increase in the average
annual number of observations (Fig.~\ref{fig:stat_year}a). 
Organization of many new stations in the early 1980s in the framework of the NASA CDP program, especially in 1982--1984, is reflected as a jump
in the number of stations in Fig.~\ref{fig:stat_cumul}e.
However, many of these stations were temporary one and subsequently did not provide long-term series of observations.
The jumps in the number of sources (Fig.~\ref{fig:stat_cumul}f) are caused mainly by special observing programs VLBA Calibrator Survey (VCS) at the VLBA 
network\footnote{https://science.nrao.edu/facilities/vlba} (\citet{Gordon2016} and references therein).

\begin{figure*}
\centering
\resizebox{\textwidth}{!}{\includegraphics[clip]{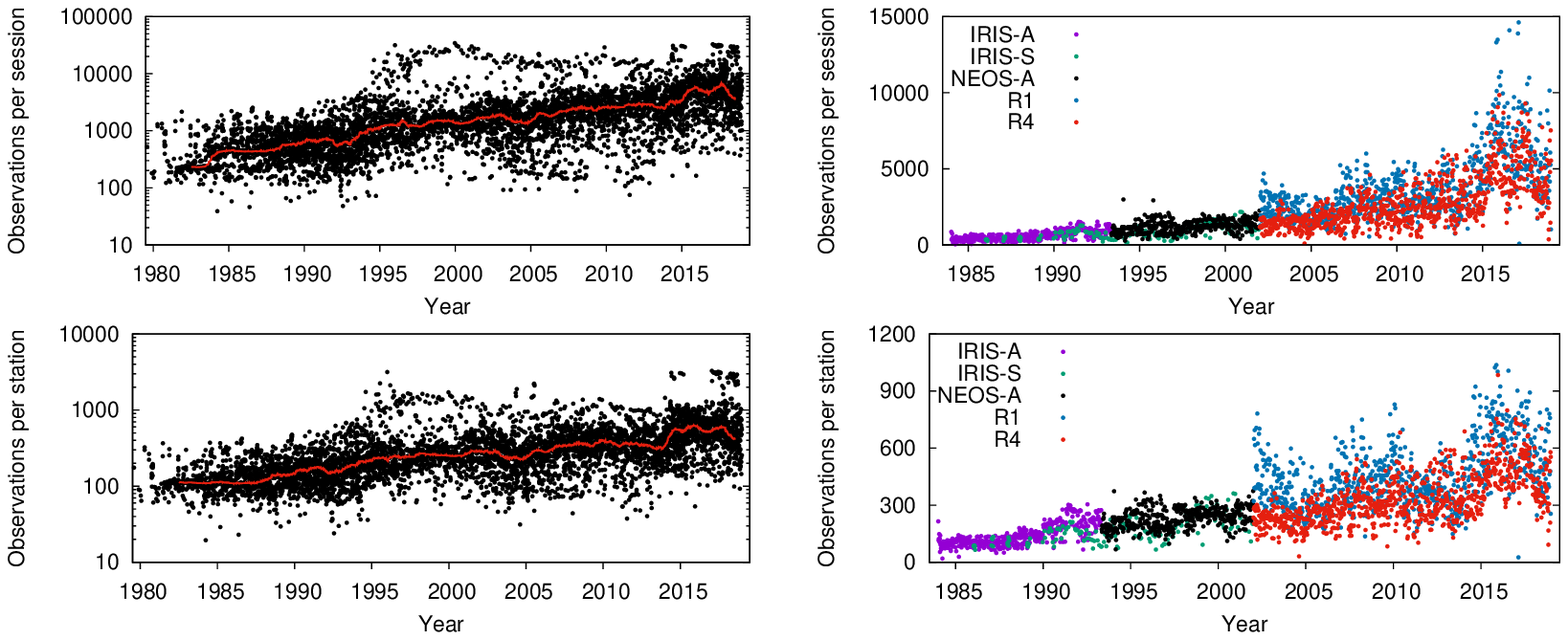}}
\caption{Number of observations per session (top panels) and per station (bottom panels): on the left, all the data; on the right, the sessions of the main programs for EOP determination.
  The red lines on the left-hand plots represent median-filtered data with a window width of 180 days.}
\label{fig:nobs_sta}
\end{figure*}

As for the average annual statistics presented in Fig.~\ref{fig:stat_year}, it is interesting to note that some characteristics
of observations grow over time, such as the number of observations and the average number of observations in one session,
while others remain nearly constant or show fluctuations around some average level over the past thirty years. 
The annual average data on the number of observations per station is detailed in Fig.~\ref{fig:nobs_sta}.
In general, one can probably say that the existing network of IVS stations of the Mark-4 standard is close to saturation
due to limited funding and resources of stations and correlators. 
The VGOS project is aimed at solving this problem.

It is interesting to trace the time required to collect each million observations (Table~\ref{tab:millions}).
One can see that while it took nearly 12 years to accumulate the first million observations, recently the number of observations in the IVS archive
has been growing, on average, by more than a million annually, especially during the years with CONT campaigns.

\begin{table*}
\begin{center}
\caption{Time required to accumulate each million observations.}
\end{center}
\label{tab:millions}
\tabcolsep=4.3pt
\begin{tabular}{|l|c|c|c|c|c|c|c|c|c|}
\hline
Million         & 1      & 2      & 3      & 4      & 5      & 6      & 7      & 8      & 9 \\
\hline
Begin           & 1979.6 & 1991.5 & 1995.4 & 1997.8 & 2000.4 & 2002.9 & 2005.4 & 2007.0 & 2008.8 \\
End             & 1991.5 & 1995.4 & 1997.8 & 2000.4 & 2002.9 & 2005.4 & 2007.0 & 2008.8 & 2010.9 \\
Number of years & 11.9   & 3.9    & 2.4    & 2.6    & 2.5    & 2.5    & 1.6    & 1.8    & 2.1    \\
\hline
\end{tabular}
\begin{tabular}{|l|c|c|c|c|c|c|c|c|c}
\hline
Million         & 10     & 11     & 12     & 13     & 14     & 15     & 16     & 17 \\
\hline
Begin           & 2010.9 & 2012.5 & 2014.1 & 2014.9 & 2015.7 & 2016.7 & 2017.4 & 2017.9 \\
End             & 2012.5 & 2014.1 & 2014.9 & 2015.7 & 2016.7 & 2017.4 & 2017.9 & 2018.9 \\
Number of years & 1.6    & 1.6    & 0.8    & 0.8    & 1.0    & 0.7    & 0.5    & 1.0    \\
\hline
\end{tabular}
\end{table*}

\begin{figure*}
\centering
\resizebox{\textwidth}{!}{\includegraphics[clip]{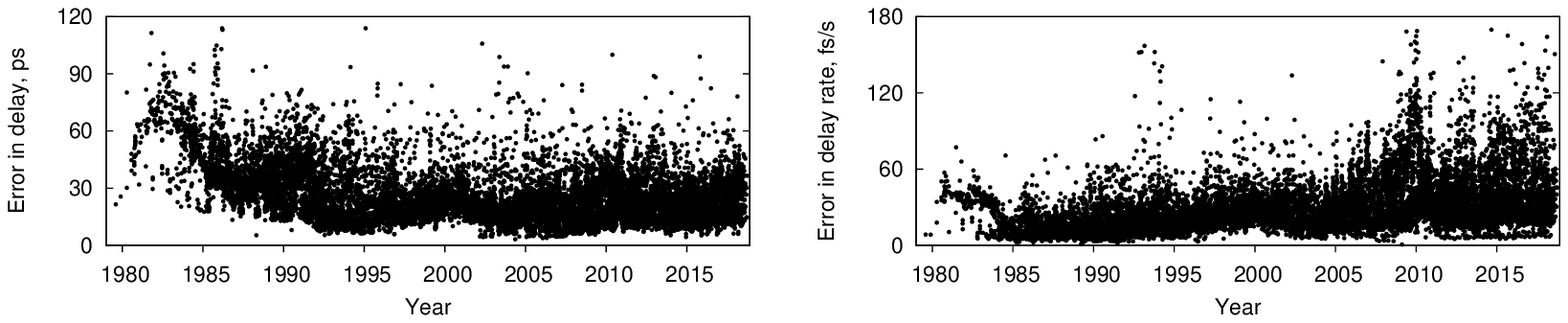}}
\caption{Error of VLBI delay (left, ps) and delay rate (right, fs/s).}
\label{fig:delay_error}
\end{figure*}

The time variation in the delay and delay rate errors reported by correlator is shown in Fig.~\ref{fig:delay_error}. 
It can be noted that the delay rate error slightly increases with time, which can be explained, first of all, by the tendency to decrease the scan duration.
Increasing the sensitivity of radio interferometric systems allows achieving a given signal-to-noise ratio in less signal accumulation time, which makes it possible to increase the number
of observations (see Fig.~\ref{fig:nobs_sta}). 
It should be noted that most scientific results are obtained using VLBI delays; however, delay rates are also used for some studies, for example, for computation
of the radio sources coordinates (\citet{Jacobs2014}; Charlot {\it et al.}, in preparation).

The above data clearly show an increase in the average number of stations participated in one 24-hour session (Fig.~\ref{fig:stat_year}h). 
The average number of observations per station per 24-hour session also increased (Fig.~\ref{fig:nobs_sta}). 
These factors made a significant contribution to improving the EOP accuracy over the past years (see Section~\ref{sect:eop}).

The maximum number of stations in one session was 32 for session 091118XA (IYA09), which was specifically organized on the occasion of the International Year of Astronomy 2009. 
The same session had the largest number of baselines with correlated data (493). 
The largest number of stations in regular observing programs was 21 (in session 131001XH) and 20 (in 10 sessions), while the maximum number of baselines was 188 (in 6 sessions).

The longest baseline included in the schedule was HART15M (South Africa) to KOKEE (Hawaii, USA), 12'723~km long. 
The longest baselines with obtained observables were SESHAN25 (China) to TIGOCONC (Chile), 12'660~km long; HN-VLBA (northeastern USA) to YARRA12M (Australia), 12'632~km long;
and MEDICINA (Italy) to WARK12M (New Zealand), 12'626~km long. 

The largest number of radio sources (370) were observed in session 150317XC (VCS-II-I), and the largest number of observations (34'221) were obtained in session 991220XA.


\section{Celestial reference frame}
\label{sect:crf}

The radio source position catalogs are one of the main and unique results of radio astrometry observations.
The idea of using extragalactic objects for maintenance of a celestial coordinate system has been discussed, according to some sources,
since Pierre-Simon Laplace and William Herschel \citep{Ma1989}. 
Of course, at that time, it was a matter of optical observations. 
The advantage of constructing a celestial coordinate system based on extragalactic objects is that the latter have no noticeable proper motions 
at the accuracy level of ground-based optical observations. 
However, although the accuracy of modern radio astrometry and optical space astrometry is sufficient to detect the displacements of extragalactic
objects at the sub-mas level, the celestial coordinate system based on those remains the basis for astronomy and geodesy.

The VLBI potential in establishing a celestial coordinate system that would be largely free from errors of proper motions
and large-scale systematic errors was quickly recognized by astrometry scientists. 
Soon after, the first proposals were made for the possible replacement of the celestial coordinate system defined by the fundamental catalog 
of star positions (FK4 at that time) with a system based at VLBI observations of extragalactic radio sources \citep{Fricke1970}. 
Finally, such a replacement was made at the 23rd IAU General Assembly in 1997, which recommended the transition from 
the FK5 optical system to the radio International Celestial Reference System (ICRS) and its implementation, the
International Celestial Reference Frame (ICRF), starting from January 1, 1998.

The first dedicated VLBI experiments in June 1969 made it possible to obtain coordinates of radio sources with an accuracy of $1-3''$ \citep{Cohen1971}. 
By the mid-1970s, the accuracy of determining the coordinates of celestial objects by the VLBI method was comparable
with the accuracy of the best optical catalogs obtained from ground-based observations ($\approx 0.1''$) and soon surpassed it 
\citep{Rogers1973,Clark1976,Fanselow1981}. 
A review of the first catalogs of radio source coordinates is given, for example, in \citet{Umarbayeva1976,Blinov1983,Walter1989} 
and papers referenced therein.

\begin{figure}
\centering
\resizebox{\textwidth}{!}{\includegraphics[clip]{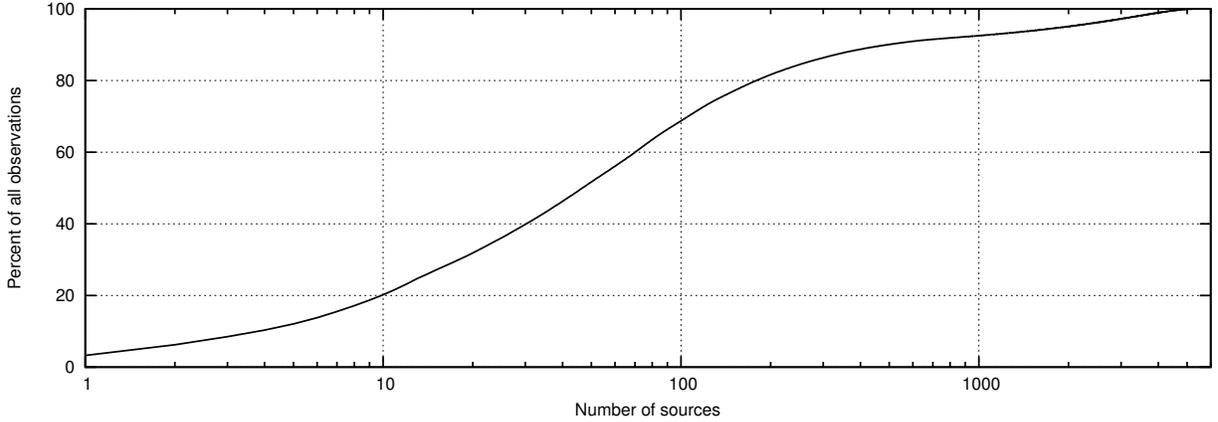}}
\caption{Percentage of the total number of observations as a function of the number of sources.}
\label{fig:nobs_sou_percent}
\end{figure}

\begin{figure*}
\centering
\resizebox{\textwidth}{!}{\includegraphics[clip]{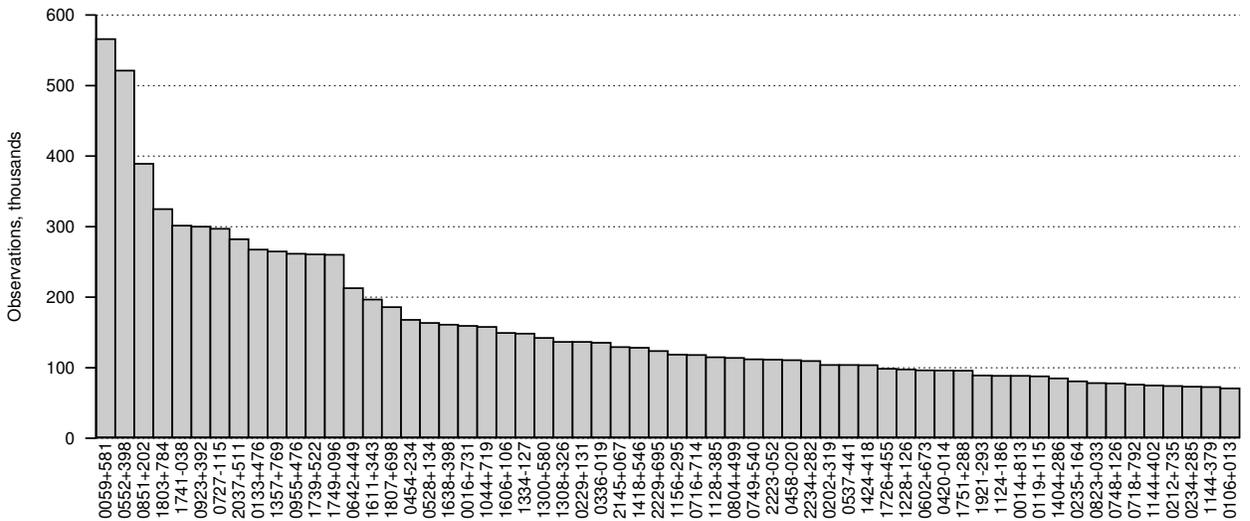}}
\caption{Number of observations (in thousands) for 60 most frequently observed sources.}
\label{fig:sou_obs}
\end{figure*}

In total, the data files of the IVS archive for 1979--2018 contain 5493 radio sources included in the observing programs.
However, for 645 of those, no good observations (suitable for use in the final processing) were obtained.
For 34 sources, only a single good observation was obtained. 
That being said, 10 sources provided more than 20\% of all observations, 47 sources provided more than a half of all observations,
and 494 sources gave more than 90\% of observations (Fig.~\ref{fig:nobs_sou_percent}). 
Figure~\ref{fig:sou_obs} shows the number of observations for the 60 most actively observed sources.
At least 1000 observations were obtained for 646 sources, and at least 100 observations were obtained for 4844 sources. 
As for good observations, more than a half of them are provided by 45 sources, and more than 90\%, by 362 sources. 
At least 1000 good observations were obtained for 568 sources, at least 100 for 4209 sources, and at least 30 good observations for 4622 sources.

\begin{figure*}
\centering
\resizebox{\textwidth}{!}{\includegraphics[clip]{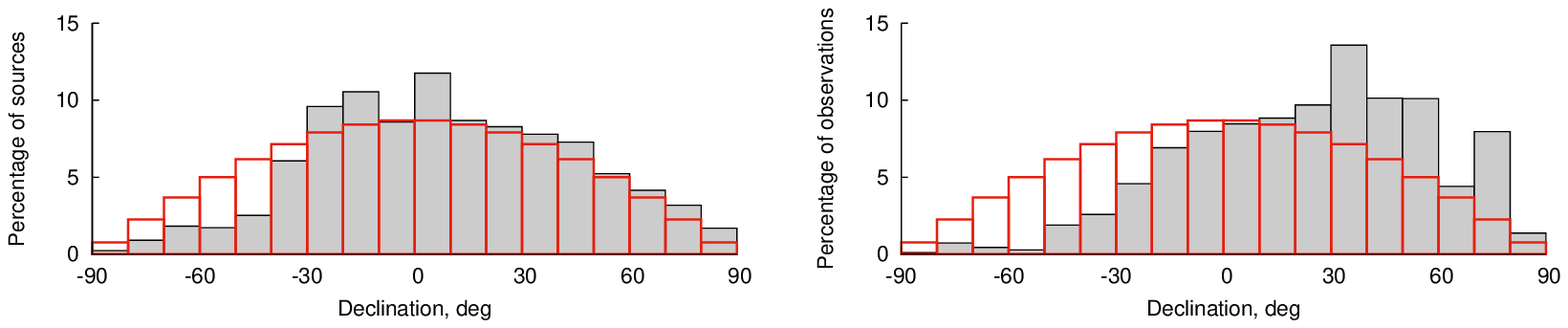}}
\caption{Distribution of the number of observations (left) and the number of sources (right) over declination as a percentage of the total number
  of observations and sources, respectively. The symmetrical stepped red line corresponds to the theoretical uniform distribution.}
\label{fig:sou_band}
\end{figure*}

\begin{figure*}
\centering
\resizebox{\textwidth}{!}{\includegraphics[clip]{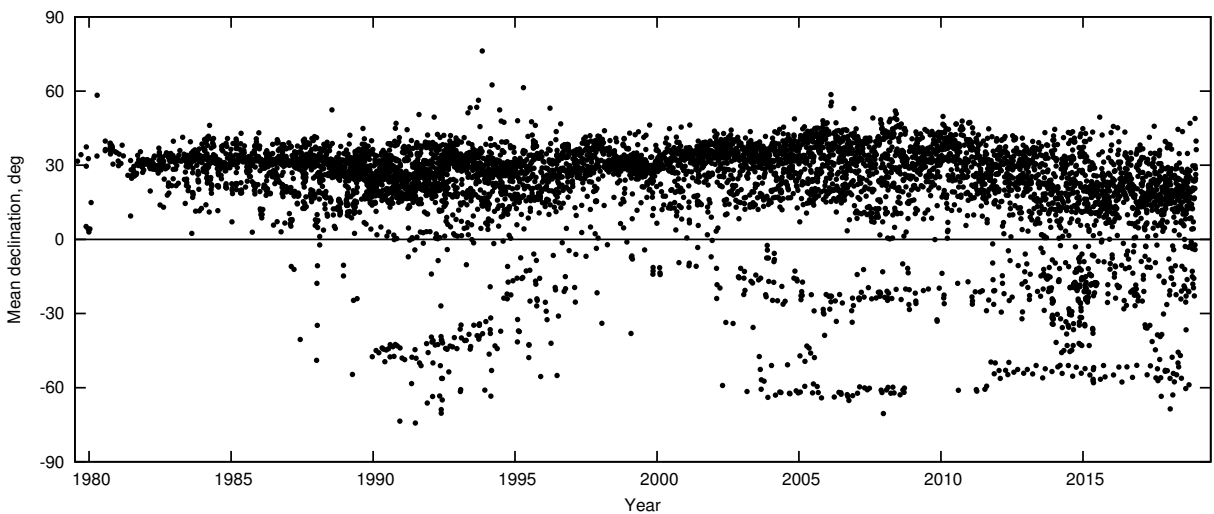}}
\caption{Average declination of sources in each 24-hour session.}
\label{fig:des_mean}
\end{figure*}

The  quality of radio source position catalogs is determined not only by the number of included radio sources, but also by the accuracy 
of their coordinates.
The uniformity of the source distribution over the celestial sphere and the uniformity of the position error distribution 
over the celestial sphere are also important. 
Both are not fully inherent in the ICRF system. 
Figure~\ref{fig:sou_band} shows that the distribution of observations over declination zones remains substantially asymmetric, 
despite great efforts to increase observational activity in the southern hemisphere of the Earth. 
These data are also confirmed by the distribution of the mean declination of sources in each 24-hour sessions (Fig.~\ref{fig:des_mean}).
It should be noted that the near absence of southern stations (there are only two relatively rarely observing stations in Antarctica
to the south of $-43^{\circ}$) leads to the necessity of observing southern sources at large zenith distances.
This leads to more sensitivity of the results by determining their declination error when considering the influence of the tropospheric delay,
especially its zenith asymmetry. 
Note that there are not only subjective, but also objective reasons for the relatively small number of VLBI stations in the southern hemisphere, 
simply due to the small area and geodetically unfavorable configuration of the continents in the southern hemisphere as compared to the northern 
hemisphere.

As for the latitudinal asymmetry of the error distribution in the radio source coordinates, the corresponding data are presented 
in Table~\ref{tab:sou_pos_error} for catalogs of the ICRF series \citep{Ma1998,Fey2015,Jacobs2014}.

\begin{table*}
\centering
\caption{Median error in the radio source coordinates for the catalogs of the ICRF series in right ascension ($\varepsilon_{\alpha*} = \varepsilon_\alpha\cos\delta$) 
  and declination ($\varepsilon_{\delta}$) for the entire catalog and for 30-degree declination zones (mas). For each catalog, the radio band is specified.
  The number of sources used is specified under the catalog name.}
\label{tab:sou_pos_error}
\begin{tabular}{c|cc|cc|cc|cc|cc}
\hline
Declination & \multicolumn{2}{c|}{ICRF (S/X)} & \multicolumn{2}{c|}{ICRF2 (S/X)} & \multicolumn{2}{c|}{ICRF3-S/X} & \multicolumn{2}{c|}{ICRF3-K} & \multicolumn{2}{c}{ICRF3-X/Ka} \\
& \multicolumn{2}{c|}{608} & \multicolumn{2}{c|}{3414} & \multicolumn{2}{c|}{4536} & \multicolumn{2}{c|}{824} & \multicolumn{2}{c}{678} \\
& $\varepsilon_{\alpha*}$ & $\varepsilon_{\delta}$ & $\varepsilon_{\alpha*}$ & $\varepsilon_{\delta}$ & $\varepsilon_{\alpha*}$ & $\varepsilon_{\delta}$ & $\varepsilon_{\alpha*}$ & $\varepsilon_{\delta}$& $\varepsilon_{\alpha*}$ & $\varepsilon_{\delta}$ \\
--90...+90  & 0.493 & 0.570 & 0.397 & 0.739 & 0.127 & 0.218 & 0.074 & 0.136 & 0.076 & 0.104 \\[1ex]
--90...--60 & 0.735 & 0.950 & 0.297 & 0.362 & 0.204 & 0.267 & 0.315 & 0.344 & 0.138 & 0.162 \\
--60...--30 & 1.358 & 1.640 & 0.586 & 1.360 & 0.176 & 0.440 & 0.172 & 0.401 & 0.129 & 0.174 \\
--30...0    & 0.723 & 0.820 & 0.417 & 0.958 & 0.132 & 0.297 & 0.071 & 0.172 & 0.079 & 0.106 \\
   0...30   & 0.372 & 0.490 & 0.386 & 0.716 & 0.112 & 0.203 & 0.064 & 0.114 & 0.061 & 0.092 \\
  30...60   & 0.373 & 0.435 & 0.336 & 0.548 & 0.114 & 0.152 & 0.067 & 0.089 & 0.057 & 0.087 \\
  60...90   & 0.332 & 0.335 & 0.546 & 0.706 & 0.144 & 0.136 & 0.061 & 0.076 & 0.058 & 0.062 \\
\hline
& \multicolumn{6}{c|}{Common sources for S/X catalogs} \\
& \multicolumn{2}{c|}{ICRF} & \multicolumn{2}{c|}{ICRF2} & \multicolumn{2}{c|}{ICRF3-S/X} & \multicolumn{4}{c}{~} \\
& \multicolumn{2}{c|}{601} & \multicolumn{2}{c|}{601} & \multicolumn{2}{c|}{601}  & \multicolumn{4}{c}{~} \\
--90...+90  & 0.482 & 0.560 & 0.065 & 0.087 & 0.040 & 0.046 \\[1ex]
--90...--60 & 0.716 & 0.920 & 0.158 & 0.216 & 0.099 & 0.112 \\
--60...--30 & 1.244 & 1.620 & 0.153 & 0.190 & 0.083 & 0.112 \\
--30...0    & 0.715 & 0.795 & 0.065 & 0.091 & 0.037 & 0.044 \\
   0...30   & 0.372 & 0.490 & 0.058 & 0.076 & 0.036 & 0.043 \\
  30...60   & 0.373 & 0.430 & 0.056 & 0.065 & 0.038 & 0.043 \\
  60...90   & 0.324 & 0.330 & 0.052 & 0.054 & 0.034 & 0.034 \\
\hline
& \multicolumn{4}{c|}{~} & \multicolumn{6}{c}{Common sources for ICRF3 catalogs} \\
& \multicolumn{4}{c|}{~} & \multicolumn{2}{c|}{ICRF3-S/X} & \multicolumn{2}{c|}{ICRF3-K} & \multicolumn{2}{c}{ICRF3-X/Ka} \\
& \multicolumn{4}{c|}{~} & \multicolumn{2}{c|}{600} & \multicolumn{2}{c|}{600} & \multicolumn{2}{c}{600} \\
--90...+90  & \multicolumn{4}{c|}{~} & 0.049 & 0.064 & 0.068 & 0.132 & 0.070 & 0.100 \\[1ex]
--90...--60 & \multicolumn{4}{c|}{~} & 0.097 & 0.113 & 0.298 & 0.312 & 0.126 & 0.142 \\
--60...--30 & \multicolumn{4}{c|}{~} & 0.086 & 0.134 & 0.163 & 0.372 & 0.112 & 0.129 \\
--30...0    & \multicolumn{4}{c|}{~} & 0.061 & 0.103 & 0.068 & 0.159 & 0.078 & 0.105 \\
   0...30   & \multicolumn{4}{c|}{~} & 0.039 & 0.045 & 0.059 & 0.110 & 0.061 & 0.092 \\
  30...60   & \multicolumn{4}{c|}{~} & 0.038 & 0.043 & 0.063 & 0.083 & 0.057 & 0.087 \\
  60...90   & \multicolumn{4}{c|}{~} & 0.033 & 0.033 & 0.054 & 0.068 & 0.058 & 0.062 \\
\hline
\end{tabular}
\end{table*}

The time variation in the basic parameters of the catalogs, such as the number of sources and the error in their coordinates,
can be traced from the catalogs computed at the GSFC and the U.S. Naval Observatory (USNO). 
Seventeen GSFC catalogs (David Gordon, private communication) and 16 USNO catalogs (Alan Fey, private communication)
obtained in 1997--2018 were used for this comparison.
These catalogs are quite uniform since they were computed using the same software and similar analysis options, but catalogs computed in different
years differ significantly due to the amount of data used and the improvement of reduction models over time.
However, it seems that the second factor, mainly the refinement of the model for accounting for troposphere-induced errors, affects primarily 
the systematics of catalogs, and not the individual errors in the coordinates of the sources that are considered in this comparison. 
The use of GSFC catalogs for this analysis is especially interesting, since the catalogs of this center serve as the basis for the ICRF catalogs 
\citep{Ma1998,Fey2015,Jacobs2014} in the S/X range.

\begin{figure*}
\centering
\resizebox{\textwidth}{!}{\includegraphics[clip]{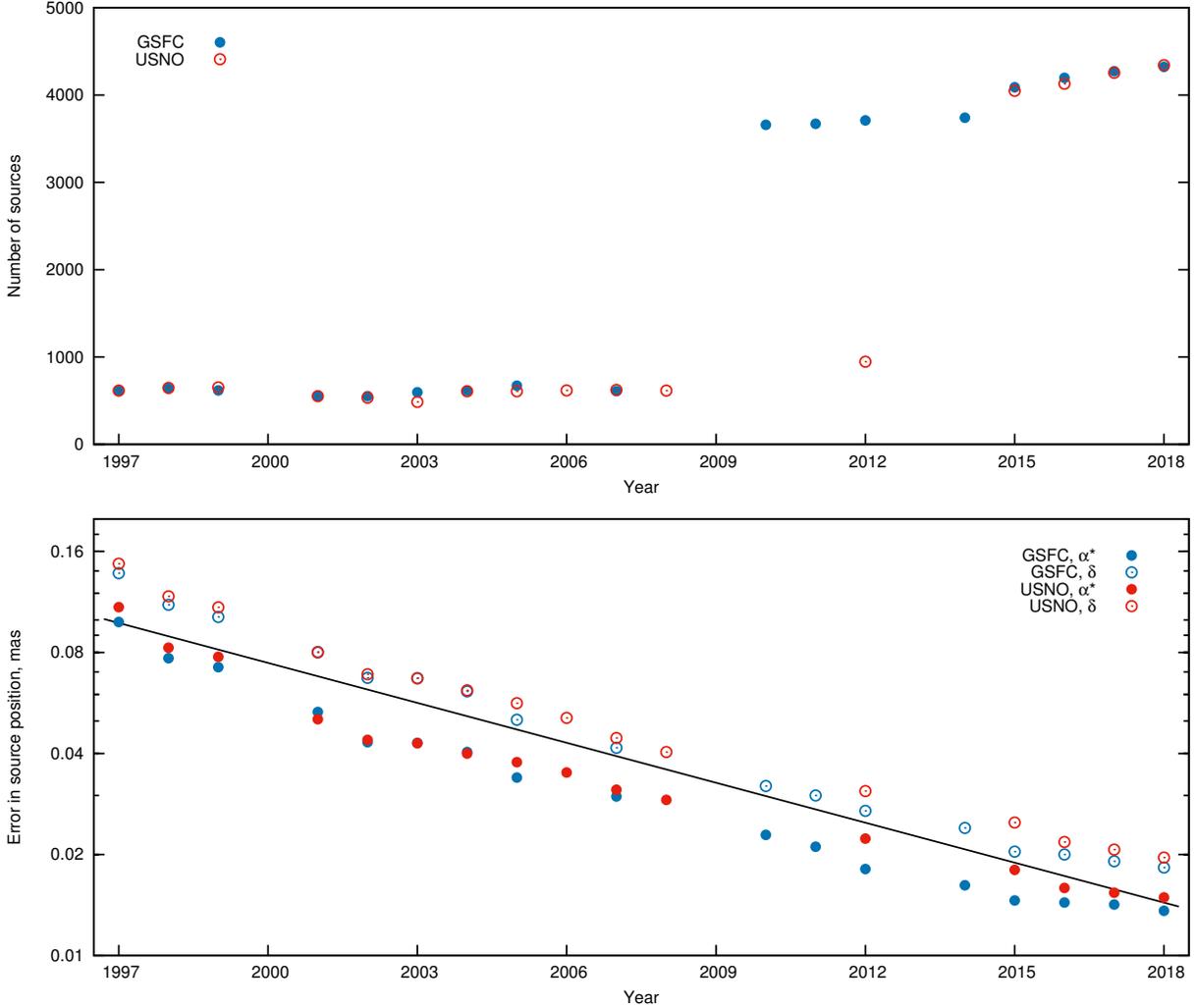}}
\caption{Number of sources (top) and median position errors (bottom) for the GSFC and USNO catalogs: solid circles show errors in right ascension;
  open circles show errors in declination. Note that the bottom graph shows the data in a logarithmic scale.}
\label{fig:err_gsf-usn}
\end{figure*}

The comparison results are illustrated in Fig.~\ref{fig:err_gsf-usn}. 
All the results turned out to be nearly the same for both centers. Since the number of sources in these catalogs
grows with time (top panel of Fig.~\ref{fig:err_gsf-usn})), the evolution of the accuracy determination of the radio source coordinates
can be reliably traced from common sources for all the 33 catalogs.  
There were 385 such sources. 
Variation in the accuracy of their coordinates over time is shown in the bottom panel of Fig.~\ref{fig:err_gsf-usn}. 
The data show that the errors in the coordinates decrease according to a power law. 
Interestingly, this is another manifestation of the power law in the results of VLBI observations, which was previously noted regarding the EOP accuracy
\citep{Malkin2009}.

In the VLBI catalogs of radio source positions, the errors in right ascension are usually smaller than the errors in declination. 
The main reason for this is the geometry of the station networks, in which most antennas are located in the northern hemisphere and,
therefore, the direction of most VLBI baselines is close to longitudinal.

\begin{figure*}
\centering
\resizebox{0.48\textwidth}{!}{\includegraphics[clip]{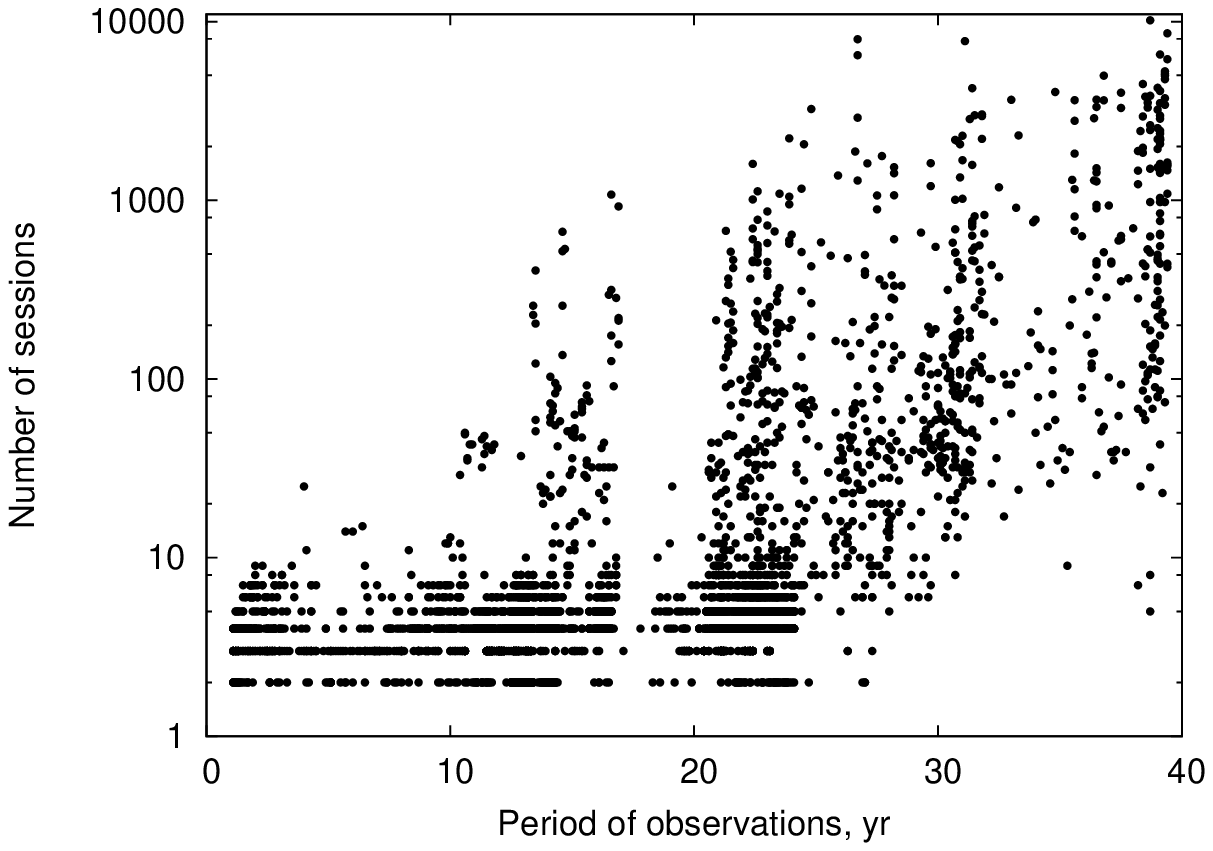}}
\hspace{1em}
\resizebox{0.48\textwidth}{!}{\includegraphics[clip]{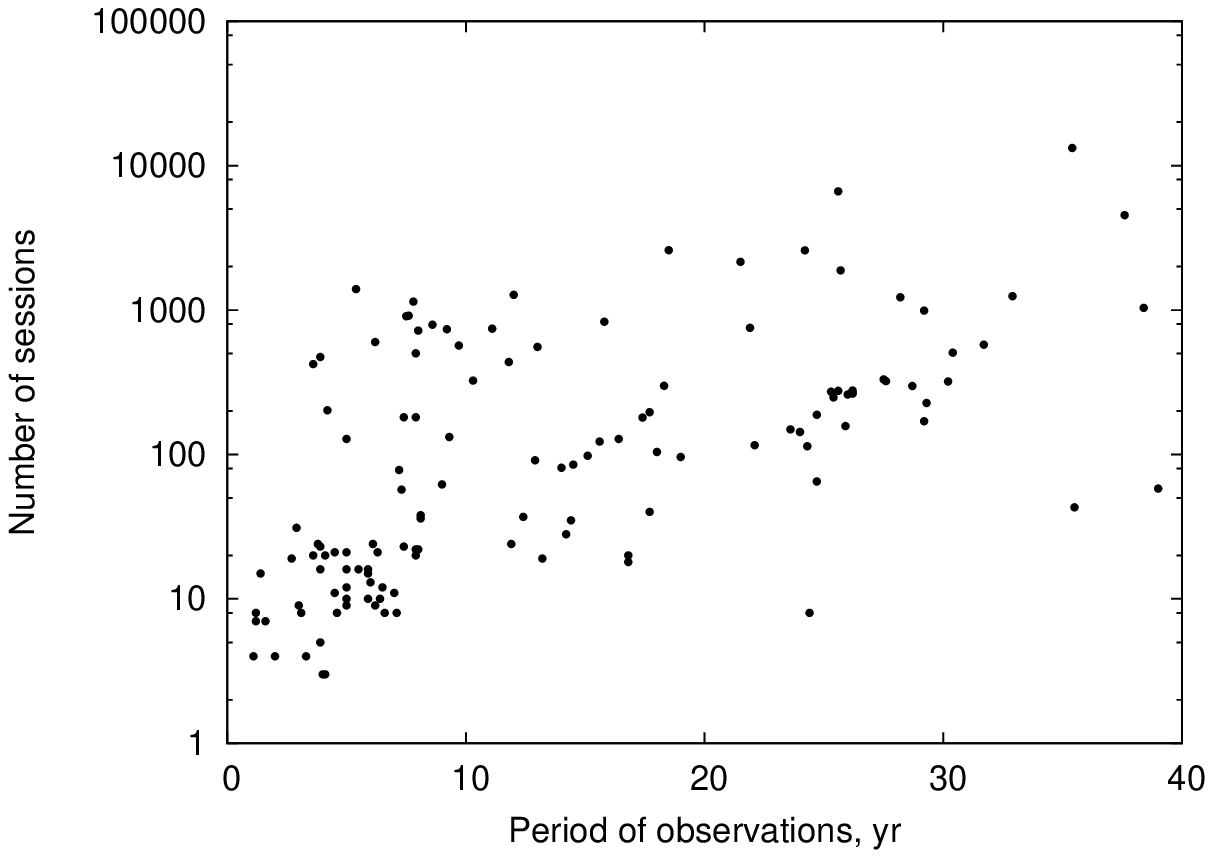}}
\caption{Correspondence between the period of observations and the number of sessions for sources (left) and stations (right).}
\label{fig:ness_span} 
\end{figure*}

In the early period of radio astrometry development, the proper motions of radio sources were considered negligible, since they are 
orders of magnitude smaller than the proper motions of stars that defined the celestial coordinate system at that time. 
However, as the accuracy of determining the coordinates of radio sources within one session increased a sub-mas level, 
their systematic displacement became quite noticeable. 
This effect becomes even more significant over time, as the accuracy of observations and analysis increases further. 
Currently, these displacements are interpreted within various models, such as the variable structure of radio sources, cosmological
models, and the galactocentric acceleration of the Solar System. 
Not all observed data can be confidently explained within the context of these models, and the reliability of these data for a particular source
depends on the number of observations and their duration. 
By the end of 2018, 4356 sources had been observed for more than five years, 4169 sources for more than 10 years, 2106 sources for more than
20 years, and 372 sources for more than 30 years. 
A long observation period does not necessarily imply a dense series of observations, which is necessary to reliably
assess the stability of the source position. 
As can be seen from Fig.~\ref{fig:ness_span} (left-hand panel), many sources with an observation period of 20, 30, or more years
have their coordinate estimates for only two or a few epochs, which is certainly not sufficient to study their apparent motion
(sources observed in one session only are not presented in the figure).

\section{Terrestrial reference frame and geodynamics}
\label{sect:trf}

The implementing VLBI in geodesy made it possible to conduct regular high-precision measurements of baseline lengths between stations,
including intercontinental baselines, which, in particular, allowed the first reliable detection of mutual movements of tectonic plates 
and regional deformations of the Earth's crust \citep{Herring1981,Herring1986,Ma1990,Ryan1998}. 
In particular, in the mid-1980s, the mutual movements of VLBI stations in California, USA, were recorded for the first time \citep{Clark1987}. 
Currently, the VLBI method plays an important role in establishing the terrestrial coordinate system, primarily, its scale and velocity field.

A total of 183 stations took part in the IVS observing programs in 1979--2018. 
Here, a station should be understood as a point on the Earth's surface with determined coordinates. 
This point can be a characteristic point of the radio antenna, usually the intersection of the axes, or a geodesic mark. 
The first is typical for permanent antennas, while the second is characteristic of mobile ones.
The combination of these points forms the terrestrial reference frame.

\begin{figure*}
\centering
\resizebox{\textwidth}{!}{\includegraphics[clip]{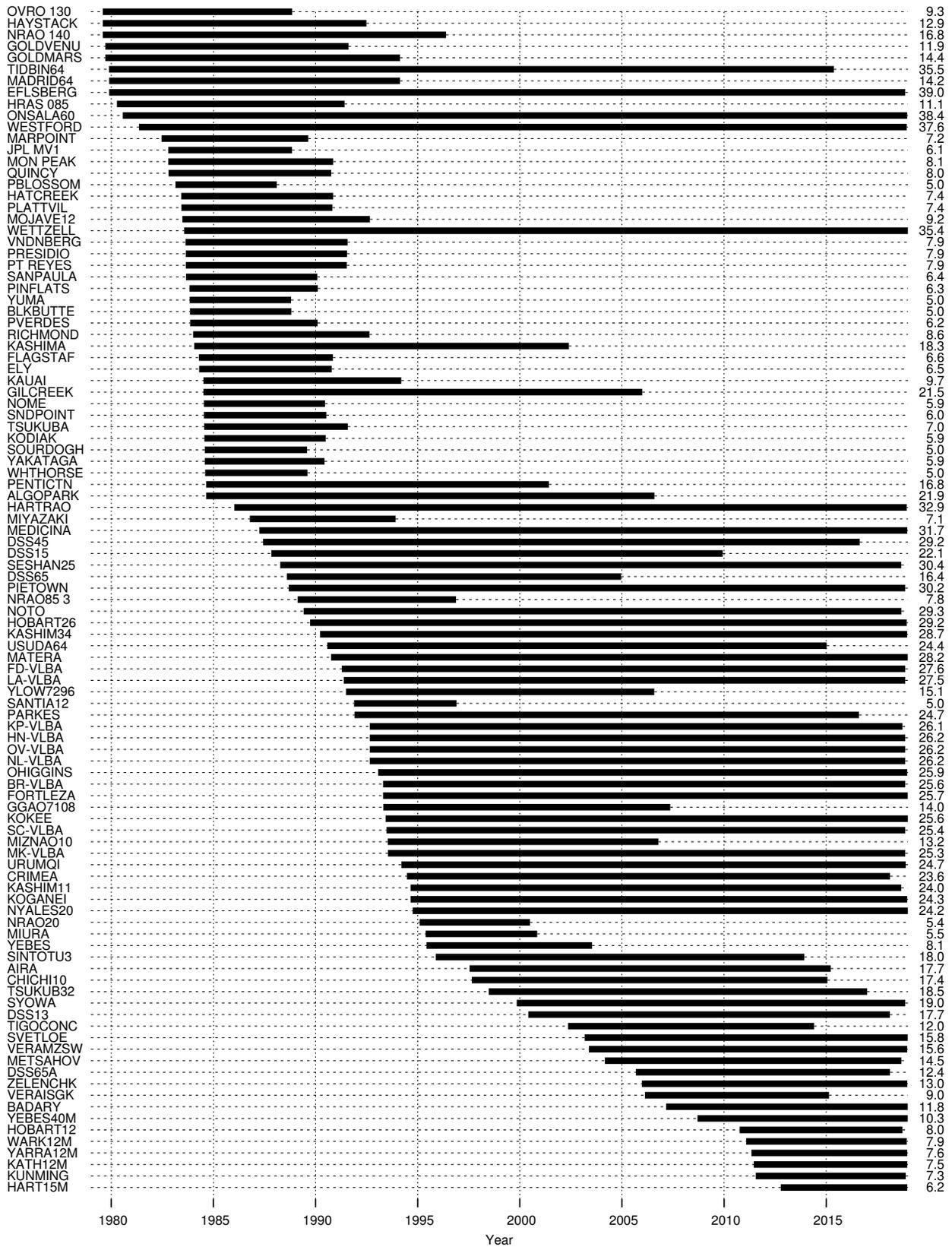}}
\caption{Operation periods of stations participated in the IVS network for at least five years. On the right is the observation period in years.
  The stations are sorted by the starting date of operation in the IVS network.}
\label{fig:sta_span}
\end{figure*}

To construct a terrestrial reference frame, it is necessary to know both the coordinates and the velocities of the stations. 
However, some of them were observed for a short time, which is not insufficient to reliably determine their velocity.
At the end of 2018, 122 stations were observed for more than 2.5 years. 
This is the minimum observation time recommended when constructing global terrestrial coordinate systems using space geodesy methods
in order to distinguish the linear trend in the station's motion (actually, its velocity) and seasonal
variations in the station's motion \citep{Blewitt2002,Altamimi2016,Abbondanza2017}.
However, VLBI stations with a shorter observation period are also included in the International Terrestrial Reference Frame (ITRF) catalog
if a station with a different observational technique has been operating there for a sufficiently long time.
Nevertheless, a longer period of observations is still desirable for a more reliable determination of the station velocity and
seasonal variations in its position, which are included in the latest ITRF realization, ITRF2014 \citet{Altamimi2016}). 
By the end of 2018, 103 stations had worked in the IVS network for more than five years, 62 stations for more than 10 years, 
36 stations for more than 20 years, and 9 stations for more than 30 years. 
The observation period for stations observing for at least five years is shown in Fig.~\ref{fig:sta_span}. 
However, the observation density is also important for reliable determination of the station velocity and the study 
of their possible nonlinear motion. 
Figure~\ref{fig:ness_span} (right-hand graph) shows that there is no direct correlation between the duration and density
of the series of station coordinates, but the correlation between these parameters is significantly higher
than for radio sources (stations that observed only in one session are not shown in the figure).

\begin{figure}
\centering
\resizebox{\textwidth}{!}{\includegraphics[clip]{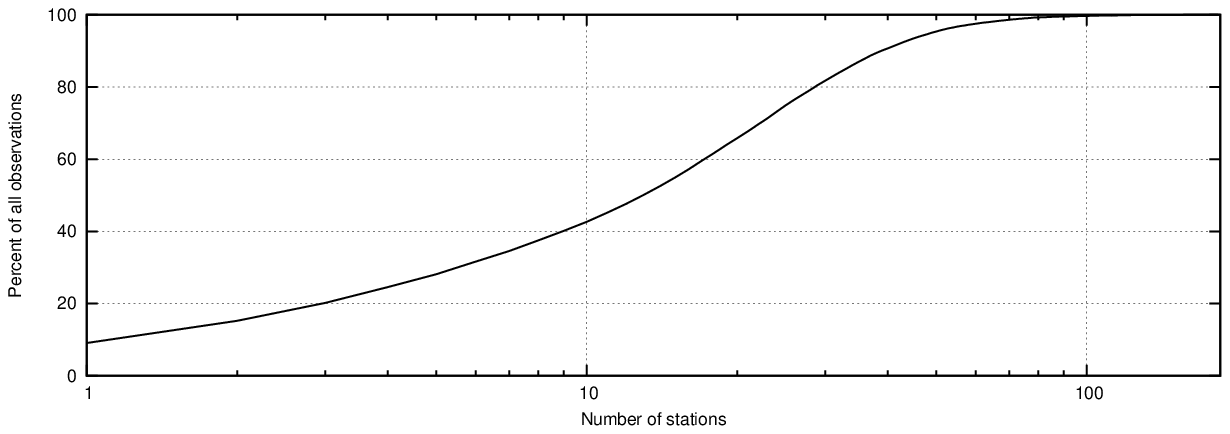}}
\caption{Percentage of the total number of observations as a function of the number of stations.}
\label{fig:nobs_sta_percent}
\end{figure}

Similarly to the case of radio sources, the total number of IVS observations is distributed between stations unevenly. 
Three stations (WETTZELL, NYALES20, and KOKEE) together provided more than 20\% of observations, 50\% of observations were made at 13 stations,
and 39 stations provided more than 90\% of observations (Fig.~\ref{fig:nobs_sta_percent}).

\begin{figure*}
\centering
\resizebox{\textwidth}{!}{\includegraphics[clip]{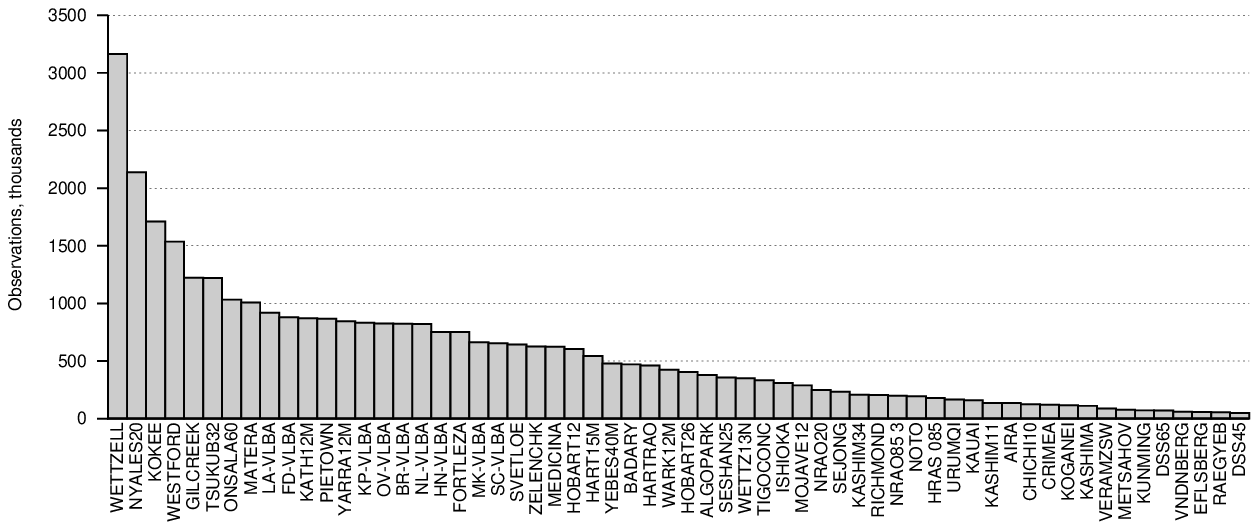}}
\caption{Number of observations (in thousands) obtained at the 60 most active stations.}
\label{fig:sta_obs}
\end{figure*}

\begin{figure*}
\centering
\resizebox{\textwidth}{!}{\includegraphics[clip]{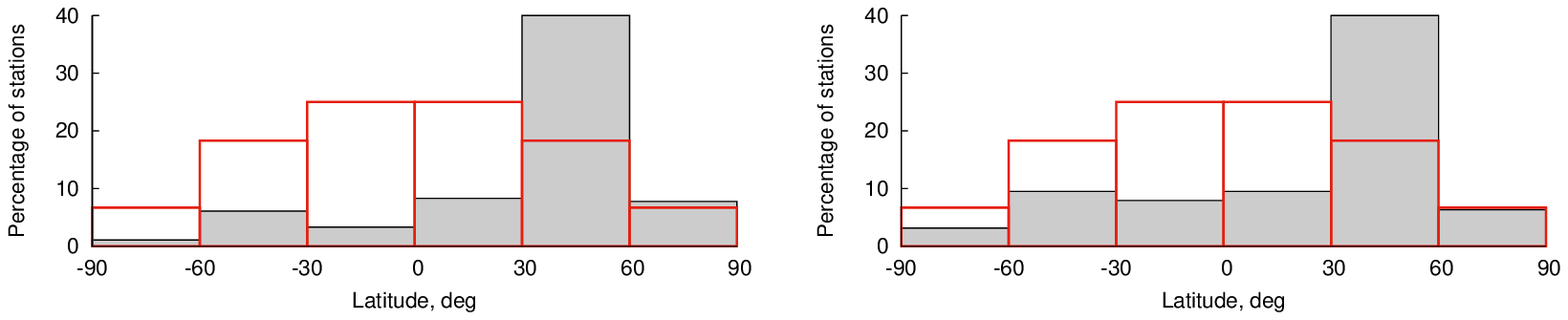}}
\caption{Distribution of stations over 30-degree latitudinal zones. On the left, all 182 stations; on the right, 63 stations observed
  at least 90 sessions. The symmetrical stepped red line corresponds to the theoretical uniform station distribution over the globe.}
\label{fig:sta_band}
\end{figure*}

Figure~\ref{fig:sta_obs} shows observational statistics for the most active stations. 
As already discussed above, the VLBI stations have an uneven distribution over the Earth's surface; most of them are located in the northern hemisphere. 
Figure~\ref{fig:sta_band} shows the distribution of the stations over 30-degree latitudinal zones.

\begin{figure*}
\centering
\resizebox{\textwidth}{!}{\includegraphics[clip]{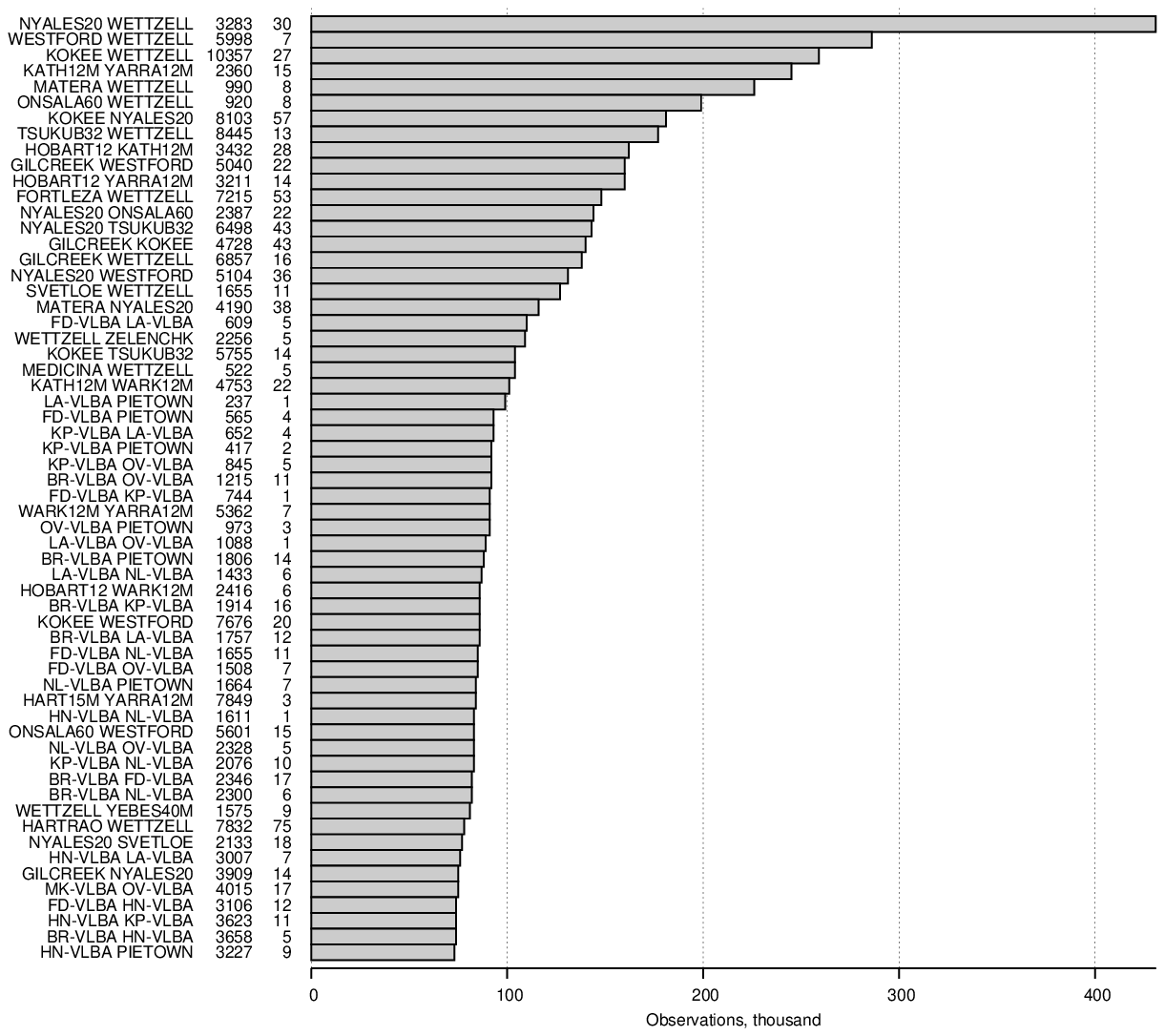}}
\caption{The number of observations (in thousands) obtained at 60 most actively used baselines between VLBI antennas. The baseline length (in kilometers)
  and the difference in latitude of the stations (in degrees) are shown after the baseline name (participating stations).}
\label{fig:bas_obs}
\end{figure*}

An important and directly observable value in VLBI is the length of the baseline between the stations, from which, in fact, the geodetic applications of VLBI began. 
Figure~\ref{fig:bas_obs} shows the statistics of the observations obtained at 60 most often observed baselines.
These statistics show that the majority of the observations were made at baselines with a predominantly longitudinal extent 
(with a small latitude difference between the stations forming the baseline), which is most likely the main reason for the higher accuracy 
of right ascensions as compared to the accuracy of declinations, as was discussed in Section~\ref{sect:crf}.

One of the main applications of VLBI is the study of variations in the coordinates of stations and the baseline lengths between them. 
It is important that the baseline length between two VLBI stations is invariant with respect to the adopted coordinate system. 
The baseline-length measurement became the first scientifically significant result of applying the VLBI method and still remains one 
of the main geodynamic results obtained by this observation method. 
Thus, the mutual movement of tectonic plates was reliably detected for the first time due to variations in the lengths of intercontinental 
baselines between American and European VLBI stations.

\begin{figure*}
\centering
\resizebox{\textwidth}{!}{\includegraphics[clip]{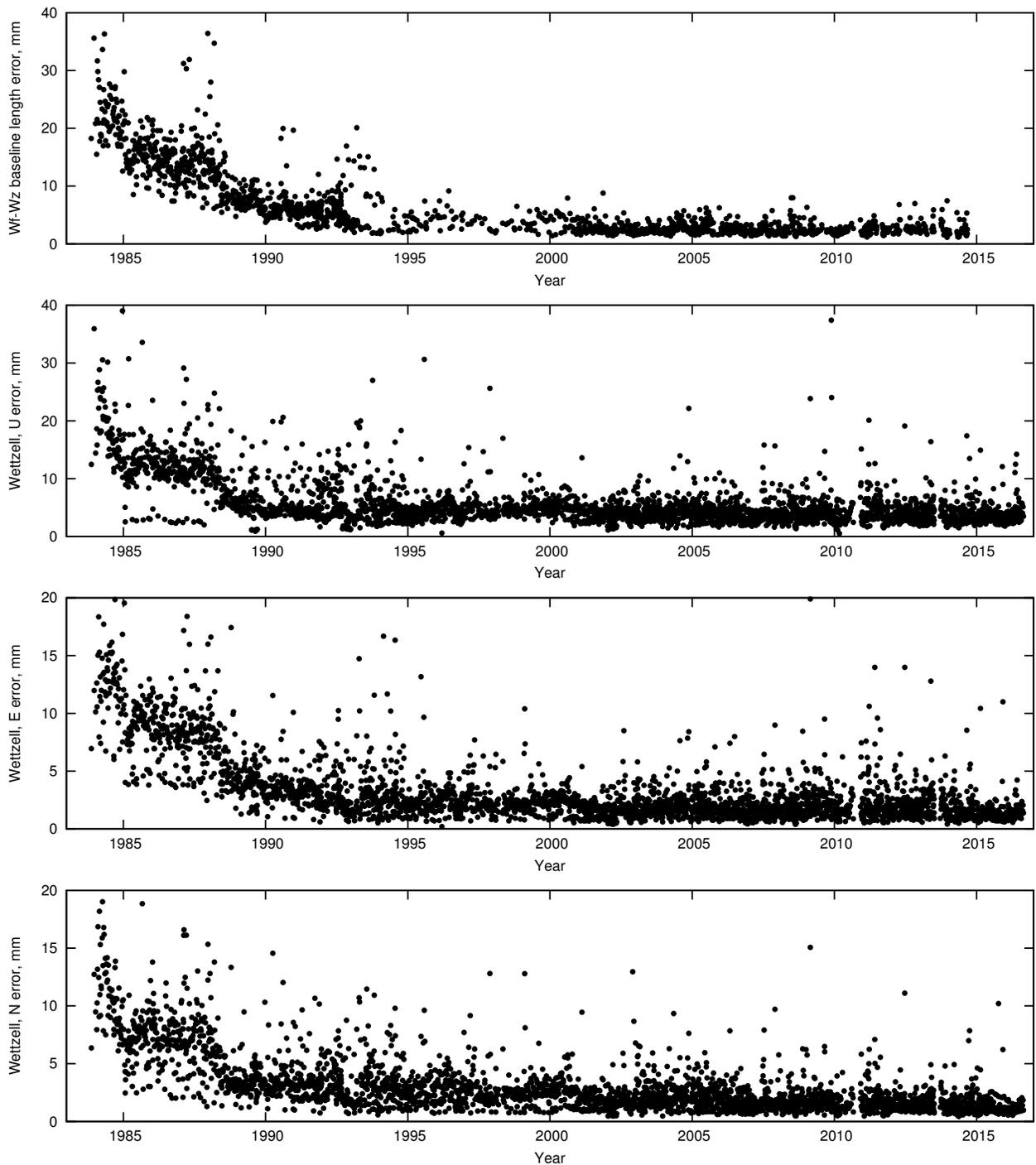}}
\caption{Accuracy evolution of baseline lengths and station coordinates over time. From top to bottom: the error in the WESTFORD--WETTZELL baseline length;
  the error in the vertical component of the position of the WETTZELL station; and the errors in the eastern and northern horizontal components of the WETTZELL station.}
\label{fig:wf-wz}
\end{figure*}

The error in determining the baseline length is traditionally used in VLBI to assess the accuracy of geodetic measurements. 
The first measurements of the baseline lengths had a meter-level accuracy; however, by the beginning of the 1970s, 
the accuracy improved to $\sim$10~cm, and by the beginning of the 1980s, it reached $\sim$2~cm \citep{Ryan1998}. 
Such a fast improvement in accuracy in the first 10--15 years of the development of VLBI is explained, first of all, by the transition of VLBI systems
from the Mark-1 to the Mark-3 standard \citep{Ryan1998}. 
A further increase in the observation accuracy occurred as a result of the transition to the Mark-4 standard.
Nowadays, the error in determining the baseline length is about 2~mm. 
The evolution of the accuracy in the coordinates of the stations and baseline lengths over the past 30 years is shown 
in Fig.~\ref{fig:wf-wz} by the example of stations WETTZELL (Germany) and WESTFORD (Eastern USA).
The data for this plot was kindly provided by Daniel MacMillan (GSFC). 
Note that the WETTZELL station observed in the largest number of sessions, and the WESTFORD--WETTZELL baseline
was the most frequently observed intercontinental baseline.

The comparison of the errors in the vertical and horizontal components of the station position shows that the errors 
of the both horizontal components are approximately the same, and the error of the vertical component is approximately twice as large, 
which is also typical of other measurements using space geodesy methods. 
After 1993, the accuracy of station coordinates and baseline length determinations almost do not improve, 
which is consistent with the accuracy evolution of the EOP discussed in the next section.


\section{Earth rotation parameters}
\label{sect:eop}

The determination of the EOP is one of the main scientific and practical applications of VLBI. 
The VLBI method is the only high-precision method that allows determining the precession and nutation of the
Earth's axis, as well as determining the Universal Time.
Without the results of VLBI, it is impossible to build modern precession-nutation theories. 
The same observations of the variations in the celestial pole position play an important role in studying the structure of
the Earth and internal processes in the Earth's body.
For this reason, EOP monitoring programs are of great importance in planning the IVS network operation. 
In 2018, out of 181 regular 24-hour IVS sessions, 105 were primarily intended to determine EOP. 
Of course, these sessions are also used for all other astrometric and geodetic solutions, but their schedule is optimized, 
first of all, to obtain the most precise EOP values. 
In addition, 366 Intensive sessions were held for rapid determination of UT1.

Until the 1980s, VLBI data were used to determine the EOP irregularly. 
A review of early efforts in determining the EOP by the VLBI method before the early 1980s can be found, for example, in \citet{Blinov1983,Moritz1987}. 
The first special observing programs for determination of the EOP were the POLARIS program of the US Geodetic Survey
(146 sessions from November 1980 to November 1990) and the JPL's TEMPO program (from the middle of 1980). 
Only the POLARIS observations are available in the IVS data center. 
In the first three years, the observations under this program were carried out at two, rarely three, US stations, and only from the end of 1983, 
the European WETTZELL station joined the POLARIS network. 
The observations of the CDP program have also been used to determine the EOP since August 1979. 

The main VLBI observing programs for the EOP determination began in 1984 and were based on observations in the Mark-4 standard. 
IRIS-A was the first of these programs; its observations were carried out from January 4, 1984 to April 26, 1993. 
At first, the IRIS-A observations were carried out once every 5 days, and once a week from the end of April 1991. 
In total, 637 sessions were obtained under this program. 
Concurrently, from April 29, 1987 to October 19, 1994, monthly observations were conducted at an alternative network of stations under the IRIS-P program. 
In total, 92 sessions were obtained under that program. 
Also, from January 9, 1986 to December 13, 2001, observations were carried out under the IRIS-S program, the network of which
included the southern station HARTRAO (South Africa) and sometimes SANTIA12 (Chile). 
Under this program, 167 sessions were obtained with an average interval of approximately a month. 
The organization of the NEOS-A program (452 sessions from May 5, 1993 to December 27, 2001), which became the main program for the EOP 
for these years, and CORE (148 sessions from January 8, 1997 to December 19, 2001), were of great importance for improving the accuracy of the EOP.
The observations of the latter, in addition to their own significance, were also used for comparison with the NEOS-A program to study
the systematic errors of the EOP determination from VLBI observations. 
Since the beginning of 2002, R1 and R4 programs have become the main IVS programs for determining the EOP. 
Their observations are conducted twice a week, normally on Mondays (R1) and Thursdays (R4) (days of the week refer to the start of the session).

\begin{figure*}
\centering
\resizebox{\textwidth}{!}{\includegraphics[clip]{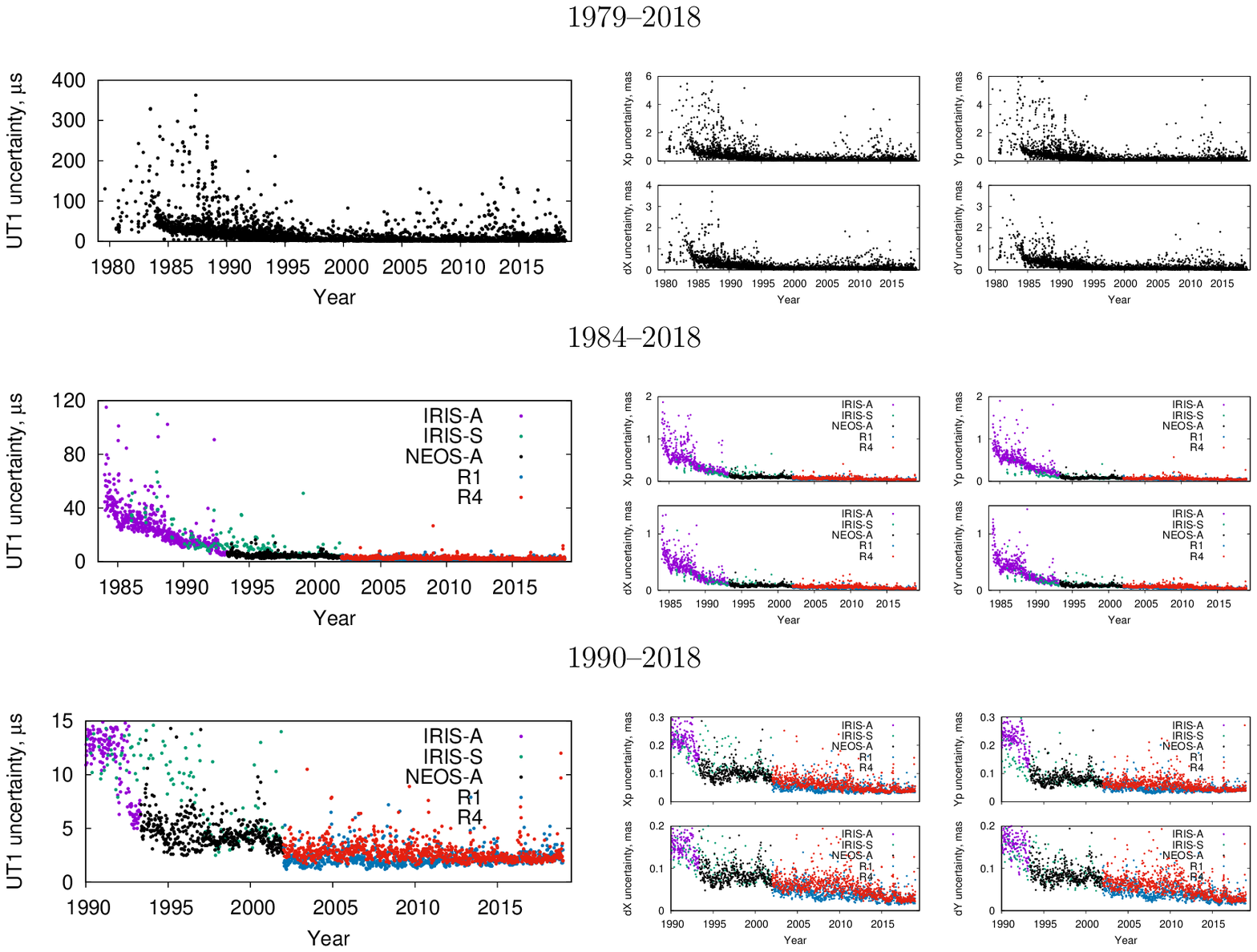}}
\caption{Time variations in the EOP errors according to the GSFC data for the entire observation period (top panel), and for the five main EOP programs
  since 1985 (middle panel) and since 1990 (bottom panel). The left panel shows the data for UT1 in $\mu$s. The time variations in the errors of the other four EOP
  (the coordinates of the pole $X_p$ and $Y_p$ and the celestial pole offsets $dX$ and $dY$) show similar behavior and are shown in the right-hand panel in mas.
  The periods of action of the main observing programs for determining the EOP: IRIS-A 1984.0--1993.0; IRIS-S 1986.0--2002.0; NEOS-A 1993.3--2002.0; R1 and R4 since 2002.0.}
\label{fig:eop_error}
\end{figure*}

The time variation in the EOP errors from VLBI observations is shown in Fig.~\ref{fig:eop_error}.
Each point in the graphs corresponds to one 24-hour series of observations. 
The data taken from an GSFC EOP series\footnote{ftp://cddis.nasa.gov/vlbi/ivsproducts/eops/gsf2016a.eops.gz} stored in the IVS Data Center. 
The same figure shows the data for the five main observing programs for determining EOP: IRIS-A, IRIS-S, NEOS-A, R1, and R4 extracted
from the same GSFC EOP series. 
The plots clearly show several points of a sharp improvement in the EOP accuracy. 
The most noticeable are the sharp increases in May 1993 after the launch of the NEOS-A program and in early 2002 after 
the transition to the R1 and R4 programs.
The features of the EOP series are important to note when determining the optimal data interval for studying
the long-period variations of the Earth's rotation,for example, to refine the precession-nutation model \citep{Malkin2014}. 

\begin{figure*}
\centering
\resizebox{\textwidth}{!}{\includegraphics[clip]{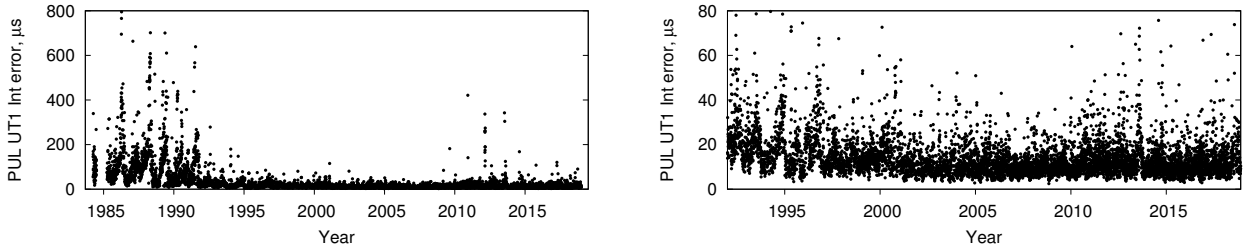}}
\caption{Time variations in the errors of UT1 (in $\mu$s) obtained from Intensive sessions as computed at the Pulkovo Observatory by the author:
  all data (left) and data after 1992 (right).}
\label{fig:int_error}
\end{figure*}

Figure~\ref{fig:int_error} shows the time variation in the UT1 errors from the Intensive sessions for the Pulkovo Observatory
series\footnote{ftp://cddis.nasa.gov/vlbi/ivsproducts/eopi/pul2010a.eopi.gz}.
Each point in the plats corresponds to one Intensive series.
Comparison with the Intensive UT1 series computed in other centers, showed that they are all close to each other.

\begin{figure*}
\centering
\resizebox{\textwidth}{!}{\includegraphics[clip]{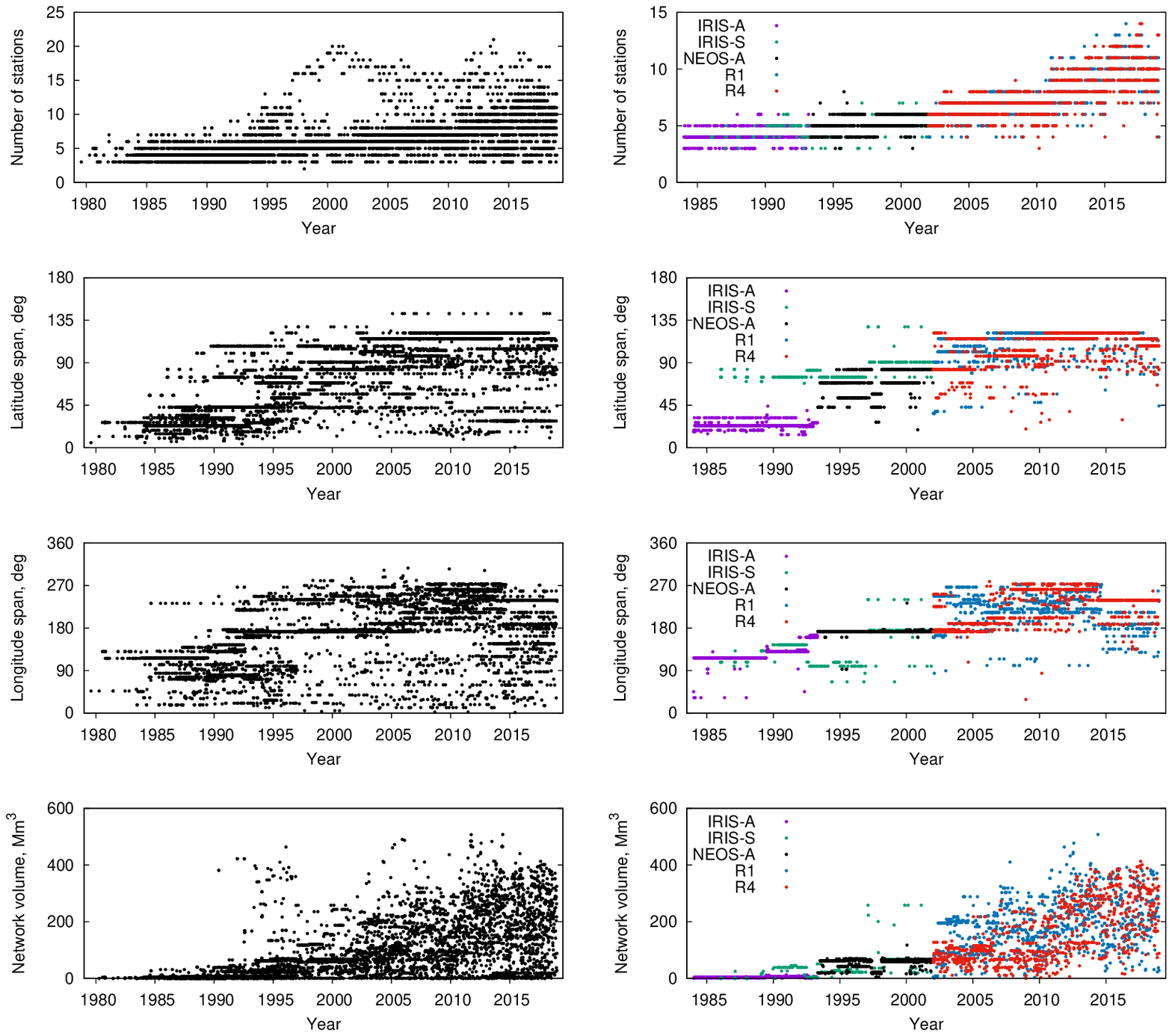}}
\caption{Time variation in the geometric characteristics of the IVS network. From top to bottom: number of stations, latitudinal network span, longitudinal network span, and network volume.
  The left panel contains the data for all sessions; the right panel contains the sessions of the main programs for EOP determination.}
\label{fig:eop_net}
\end{figure*}

As was shown in \citet{Malkin2009}, both the internal and external errors of the EOP correlate well with the size of the network, primarily 
with its geometric volume.
Figure~\ref{fig:eop_net} shows the time variations in the geometric parameters of VLBI networks, the observations from
which are collected in the IVS archive. 
A comparison of these data with the data on improving the accuracy of the EOP given above confirms their close dependence.

\section{Conclusion}
\label{sect:conclusion}

By its 20th anniversary, IVS has become the leading international organization that coordinates the work of many
institutions around the world working in the field of radio astrometry and VLBI sub-system of space geodesy.
Currently, the IVS observing network includes approximately 60 VLBI antennas located in many countries on all continents (57 antennas were involved in the observations of 2018).
The IVS also includes technology development centers, operation centers that perform planning of observations and coordinate the work of stations, data centers, and data analysis centers.

In total, tens institutions from more than twenty countries participate in the IVS operations.
The IVS data centers have so far accumulated more than 17 million observations and, on average, more than a million observations are added annually.
The results of these observations are used with high weight in the derivation of the ITRF and EOP.
The observations stored in the IVS data centers are used by the IVS in cooperation with the IAU to derive the ICRF, the official IAU celestial frame since 1998.
The VLBI is also the only high-precision modern method for determining UT1 and the celestial pole motion.
In addition, VLBI plays a unique role in maintaining the long-term stability of the EOP series and the ITRF.

Currently, the IVS continues its active development.
A network of new-generation VGOS stations based on fast antennas, broadband high-speed data acquisition and recording systems is growing fast.
In addition to the already operating stations mentioned above, new stations are being commissioned such as 13-meter telescopes
installed at the observatories SVETLOE, ZELENCHK, and BADARY of the Institute of Applied Astronomy of the Russian Academy of Sciences \citep{Ipatov2014,Nosov2019}.
New methods of planning and controlling the operation of VLBI stations are being introduced into the practice of the IVS network.
To increase the accuracy of processing the results, astronomical and geophysical models and data analysis methods are being improved.

All this makes it possible to speak confidently about the strengthening of the role of radio astrometry in astronomy, geodesy,
Earth Sciences, as well as in solving applied problems in the coming years.


\section{Acknowledgements}

All the VLBI observations and results used in this study were obtained through years of hard and highly skilled 
work by many people and organizations who contributed to the IVS activity: stations and correlators, technology
development, operation, data, and analysis centers, as well as IVS coordinating bodies.
Their contribution to the development of science deserves the highest recognition.

The author is grateful to Daniel MacMillan (NVI Inc., NASA Goddard Space Flight Center) for the series of station coordinates and baseline lengths, 
as well as to David Gordon (NVI Inc., NASA Goddard Space Flight Center) and Alan Fey (US Naval Observatory) for the catalogs of radio source positions provided for this study.

Observations for 2018 were partially processed using the vgosDB-to-NGS data converter\footnote{https://github.com/AstroLis/VGOSdb2NGS} which was kindly provided by Svetlana Mironova,
Elena Skurikhina, and Sergey Kurdubov (Institute of Applied Astronomy of the Russian Academy of Sciences).

The author thanks the anonymous reviewer for valuable comments and suggestions which helped to improve the manuscript.

This work was partly supported by the Russian Government program of Competitive Growth of Kazan Federal University.

This research has made use of the SAO/NASA Astrophysics Data System\footnote{https://ui.adsabs.harvard.edu/} (ADS).

\bibliography{ivs_history_v3}
\bibliographystyle{joge}

\end{document}